# Design and fabrication of fiber optic microlenses using an arc fusion splicing system

Szymon Matczak, Dorota Stachowiak, and Grzegorz Soboń

*Abstract*—In this study, we introduce an new approach to fabricating fiber optic microlenses using a three-electrode arc fusion splicer. Through beam propagation method-based simulations, we verified the performance of our lenses, achieving highly consistent results across both simulations and experiments. We fabricated three distinctive microlens types: ball-type, fiber tip, and ball-type on a tapered fiber, demonstrating versatility in lens shape and function. By precisely adjusting lens designs, we achieved beam radii between 2.7 µm and 23.5 µm, with focal lengths spanning from 12 µm to 787 µm. Our method offers robust control over lens geometry, enabling tailored beam parameters for specific applications. This technique is efficient, cost-effective, and adaptable to various optical fibers, including multimode and polarization-maintaining, highlighting its potential for broader optical and photonic applications.

*Index Terms*—optical fibers, fiber microlenses, fiber optics, ball-type microlens, fiber tip microlens

## I. INTRODUCTION

THE dynamic growth of photonics requires the development of innovative technologies for fabricating non-standard fiber optic components. The most demanding applications, like biomedical fiber optic sensing, integrating all-fiber systems with photonic chips, manipulating nanoparticles using optical tweezers, or optical imaging [1-5] rely on the availability of customized components. An excellent example of such components is a fiber microlens formed at the very end of a fiber. Such a lens focuses light at a specific distance from the end of the fiber, eliminating the need for bulk optics. This makes the coupling system more compact and increases the efficiency of coupling light from the fiber to the waveguide or other non-standard fibers. In many fiber optic sensing devices, microlenses are commonly used to modify the size and shape of the beam. The light beam emitted directly from an optical fiber has a relatively large divergence angle, which can decrease coupling efficiency and limit the resolution of fiber optic sensors, such as in deep-hole measurements [6]. Another example of fiber microlens application can be found in optical sensing, where properly shaped microlens could collect reflected or scattered light from the tested sample [4]. Optical fiber microlenses enable the trapping of microparticles. Typically, a high numerical aperture (NA) microscope objective is required to focus the laser beam in traditional optical tweezers. Despite significant advances in various fields, the bulky structure of conventional optical tweezers limits their use in several environments. To make the tool easier and more convenient to manipulate, fiber optical tweezers have been developed. Utilizing optical fibers to trap micro-objects is much easier to handle and more suitable for practical applications such as trapping, levitating, and rotating microscopic particles [7]. Another application can be found in medicine, such as endoscopes. For example, laser Doppler vibrometry is used during middle ear examination to determine the movement of ear components [8]. Using a fiber-based probe for such measurement with properly designed microlens could improve the detection efficiency of such devices.

Microlenses at the end facets of optical fibers can be manufactured using various techniques, including photolithography, $CO_2$ laser machining, etching, mechanical fabrication, the microcompression molding method, electric arc discharge splicing, two-photon direct laser writing, and others [9-10]. Photolithography is historically the oldest method and allows the fabrication of microscopic microlenses (diameters between 4 µm and 10 µm). However, this method allowed fabrication only on the fiber core surface (the UV light propagated through the core, which ensured automatic alignment of the microlens), which limits its usability [9,11]. $CO_2$ laser micro-machining offers fiber processing in its whole geometry but requires complex equipment (a workstation with a $CO_2$ laser that requires periodic maintenance and service) [11-16]. The dynamic development of optical materials has led scientists to develop various techniques for fabricating and processing optical fibers. Researchers have started to develop methods such as etching using a hydrophobic effect, which allowed the fabrication of arrays of microlenses. Yet, it would require another technology to assemble microlens with fiber [17-19]. Another method of fabrication is two-photon direct laser writing. This method offers high precision and flexibility for micro-optical fabrication, but it comes with challenges, such as high costs and slower speeds compared to some other

This research was funded in whole by the National Science Centre (NCN) under the grant no. 2021/43/O/ST7/00757. For the purpose of Open Access, the author has applied a CC-BY public copyright license to any Author Accepted Manuscript (AAM) version arising from this submission. All authors are with Laser and Fiber Electronics Group, Faculty of Electronics, Photonics and Microsystems, Wroclaw University of Science and Technology, Wybrzeże Wyspiańskiego 27, 50-370 Wrocław, Poland. Corresponding author e-mail: grzegorz.sobon@pwr.edu.pl

methods [20]. Another technique is mechanical fabrication using a microgrinding machine [21-22]. This method allowed the production of high-quality microlenses, however, the technological process is very complicated and time-consuming, and it carries a risk of damaging the optical fiber. Another technique is the microcompression molding method with microlenses made of various polymer materials (PMMA, PC, ZNR, Topas COC) [10, 23-24]. The disadvantage of this method is that it requires attaching microlens to the fiber end facet as with etching with the use of a hydrophobic effect [17]. Another popular fabrication method is connected with telecommunication splicers. In this case, an electric discharge arc created by two electrodes allows melting fibers, thus splicing, tapering, and forming microlenses. However, the main disadvantages of this method are uneven heating and a long fabrication process time (up to several minutes). This increases the risk of geometry being crooked due mainly to the influence of gravity but also due to external factors like vibration occurring during the process. Another limitation is that telecommunication splicers have typically two electrodes, which causes non-uniform heating of the external surface of an optical fiber. This limits the ability to precisely shape structures. All of which can cause asymmetry of a structure, especially when fibers are not perfectly aligned during the arc discharge [25-26]. The boundaries of arc-fusion splicers can be overcome by laser- or filament furnace-based splicing and glass processing systems. They provide great precision and control of the fabrication process. However, laser systems are expensive and require periodic maintenance of a $CO_2$ laser by a qualified producer. On the other hand, electrode-based systems maintenance can be performed by a user, which makes laser-based systems less accessible for certain applications [27]. Filament-based systems require high-purity argon gas (99.999%), and filament elements that generate heat are sensitive to overheating and damage [28]. S. Park et al. managed to fabricate several microlenses with a diameter of around 70 μm. In this study, the authors used a fusion arc splicer and focused on the influence of beam expansion region. The recorded range of beam diameter was 13.9 – 23.5 μm, while the working distance range was 327 – 415 μm (no correlation indicated) [30]. B. Guzowski et al. performed a realization of ball-lens type microlenses using a FITEL S153A fusion splicer. Fabricated microlenses diameters ranged from 133 μm to 230 μm. The authors measured only the focal lengths of the microlenses. The noted range was 166.7 μm – 423.4 μm. Measurements also differed from simulations performed with MATLAB and OSLO software [29]. L.M. Wurster et al. demonstrated a different approach when producing the final shape of the microlens. When the microlens itself was created by a filament-based system, the authors decided to also polish the surface of a ball-type microlens. Nevertheless, they achieved diameters of around 400 μm each microlens. The measured spot size diameter range was 45 – 112 μm dependent on the microlens [31]. S. Balakrishan et al. presented work with a bunch of ball-type microlens probes fabricated with a 3-electrode arc discharge-based system. Achieved spot sizes ranged from 33 μm to 81 μm with working distances from 1 μm to 3.8 μm. In this case, an additional factor was the use of grated index fiber, which seems to extend both, spot size and working distance [28]. The latest milestone in the field of microlenses fabrication is the lab-on-fiber (LOF). In [32] F. Piccirillo et al. created a roadmap and described that this technology allows the placement of optical elements directly on the tip of an optical fiber, allowing for the creation of compact devices that are easy-to-use and plug-and-play. Manufacturing methods can evolve through LOF technology, enabling the creation of subwavelength platforms on fiber tips with high throughput. The integration of multifunctional components into the fiber end-face would mark a significant breakthrough in LOF technology.

All of the methods mentioned above are effective in fabricating optical fiber microlenses, although they are complicated and challenging to implement due to the complexity of the technological processes. In this work, we present our concept and fabrication details of various optical fiber microlenses and tips fabricated with a three-electrode fiber processing workstation (Large Diameter Splicing System, 3SAE, LDS). The structures were first modeled in the RsoftCAD BeamPROP software for an operating wavelength of 1550 nm. Thanks to the precise control over microlens geometry achieved using a three-electrode arc fusion splicer system, we successfully fabricated three types of optical fiber microlenses and tips on standard single-mode fibers. Additionally, the use of a coreless passive fiber (CPF) in the microlens and the ability to accurately adjust its length offer superior customization options compared to laser- and filament-based systems [27,28]. Thanks to the short processing time (≤1s) and uniform arc discharge we managed to minimize gravitational and vibration influence, and thus eliminate the imperfections of the structures shape [33]. The radii achieved in the focused beam profile of the fabricated microlenses vary from 2.5 to 6 μm, and the focal length from 42 μm to 787 μm, depending on the shape, with the beam ellipticity equal to 99% across all the microlens designs. This fact is a huge advantage as we can easily adjust output beam parameters to demanded applications without limitations to a single shape as it is by using conventional splicers. The experimental results were compared with simulations, where we showed relationships between the geometrical parameters of the fabricated structures and the parameters of the outgoing beam. Our results show that three-electrode discharge technology allows to make structures with a wide range of parameters and shapes with a short fabrication time, low maintenance effort, and low cost.

II. DESIGN, SIMULATIONS & FABRICATION OF MICROLENSES

Numerical simulations of designed structures were performed using BeamPROP software based on the beam propagation method (BPM) [34]. Modeled structures were simulated with a wavelength equal to 1550 nm. All investigated structures were fabricated using the LDS system. In contrast to conventional arc-fusion splicers, it possesses three electrodes for plasma discharge, which is advantageous for demanding applications. Its heat zone is isothermal around the fiber or multiple-fiber structure, creating a relatively narrow so-called Ring of Fire (ROF) [35]. Using the LDS, we could easily control parameters such as lengths of the specific sections of fibers, taper lengths, and microlens diameters. Controlling these parameters is crucial for adjusting the beam radius and the focal length of the output beam.

*A. Ball-type microlens*

The design of a ball-type microlens is presented in Fig. 1 (a). It was fabricated at the end of a coreless passive fiber spliced to a standard single-mode fiber (SMF-28e+). The microlens was designed to be formed from the CPF (FUD-3582, MM-125-FA, Coherent) due to its large numerical aperture (NA) of 0.46, which allows the light beam to propagate freely from the SMF. Furthermore, since it is a fiber without a core, the refractive index of the microlens is homogeneous, which is the refractive index of the CPF fiber ($n_{CPF}$ = 1.4507). The variable parameters during the simulations were the length of the CPF fiber ($L_{CPF}$) and the diameter of the microlens ($D$). In the model, we changed the CPF length between 100 and 1500 µm and the lens diameter between 140 and 320 µm. We found experimentally that 320 µm is the upper limit of the ball lens size due to technological limitations – a larger ball tends to bend the fiber at the end.

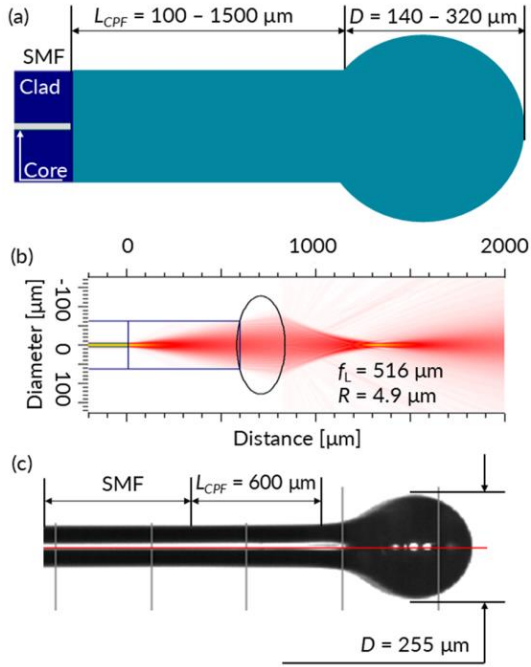

**Fig. 1.** The ball-type microlens created on the coreless fiber end face: (a) scheme modeled in the BeamPROP software, (b) example of simulated light propagation profile with focal length ($f_L$) and waist radius ($R$), (c) example of a fabricated structure with CPF length of 600 µm and ball size of 255 µm.

Fig. 1 (b) shows an example of a simulation result of a structure presented in Fig. 1 (a) with a CPF length of 600 µm and 255 µm of microlens diameter. These input parameters resulted in achieving the beam radius value equal to 4.9 µm and the focal equal to 516 µm. Fig. 1 (c) presents an LDS camera view of the fabricated structure. The microlens was formed by splicing a segment of CPF to the SMF, cleaving the CPF to the desired length, and finally molding the end of the CPF by high temperature created by the ROF in a short time (around 1 s). To create the desired microlens, the length of the CPF segment must be adequately estimated, considering how much material needs to be molded to create the demanded diameter. Based on that, the microlens diameter can be easily controlled by adjusting the arc power of the ROF.

*B. Fiber tip microlens*

The second investigated structure, shown in Fig. 2 (a), was a fiber tip made using only the stretching process. The structure was designed so that the single-mode optical fiber was tapered to achieve a maximum taper ratio TR = 125 (tip diameter equal to 1 µm). The variable parameter during the simulations was the taper length which was possible to control during the fabrication process in the range of 500 – 2000 µm.

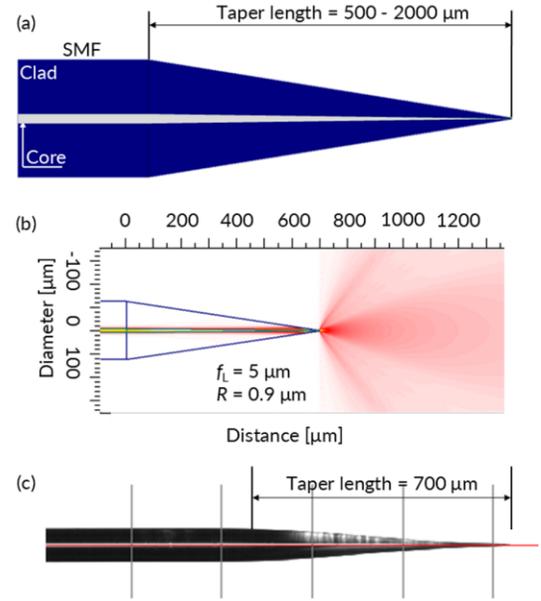

**Fig. 2.** The fiber tip microlens: (a) scheme modeled in the BeamPROP software, (b) simulated light propagation profiles with focal length ($f_L$) and radius ($R$), (c) example of a fabricated structure with taper length of 700 µm.

Fig. 2 (b) illustrates a sample simulation of the structure presented in Fig. 2 (a). The taper length was set to 700 µm and the taper ratio was set to 125. These input parameters resulted in achieving the beam radius value equal to 0.9 µm and the focal equal to 5 µm. Fig. 2 (c) presents an LDS camera view of the fabricated structure, where the taper length was equal to 700 µm. The tip was formed by subjecting the fiber to the ROF for several seconds while stretching it at a certain speed, until it broke, due to a too small waist diameter. As a result, a sharp tip was created, that could be rounded in case it is not symmetrical and the output beam's profile is not Gaussian-like. We decided to use only SMF in this case, as this approach provided flexibility in creating the fiber tip by just tapering the SMF directly, achieving a sharp tip without any extra need for the refractive index calculation or beam propagation connected with CPF.

## C. Ball-type microlens on a tapered fiber

The third designed structure was a microlens formed of the tapered section at the end of the SMF. The structure model includes a ball created of the core inside the ball created from the cladding. The variable parameters during simulations were the diameter of the tapered section ($D_{TAPER}$) and the diameter of the microlens ($D$), as shown in Fig 3 (a).

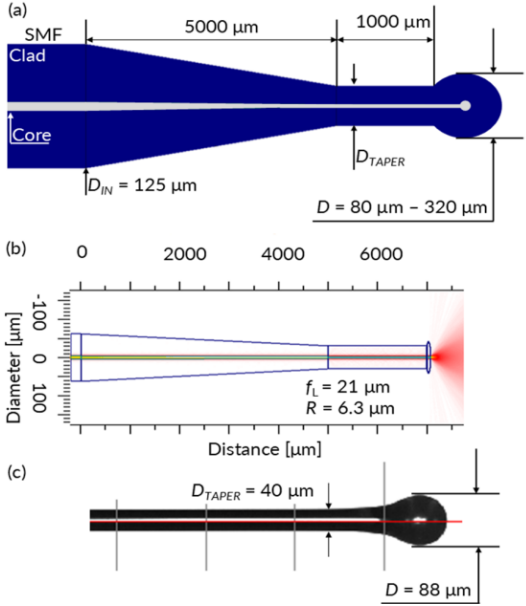

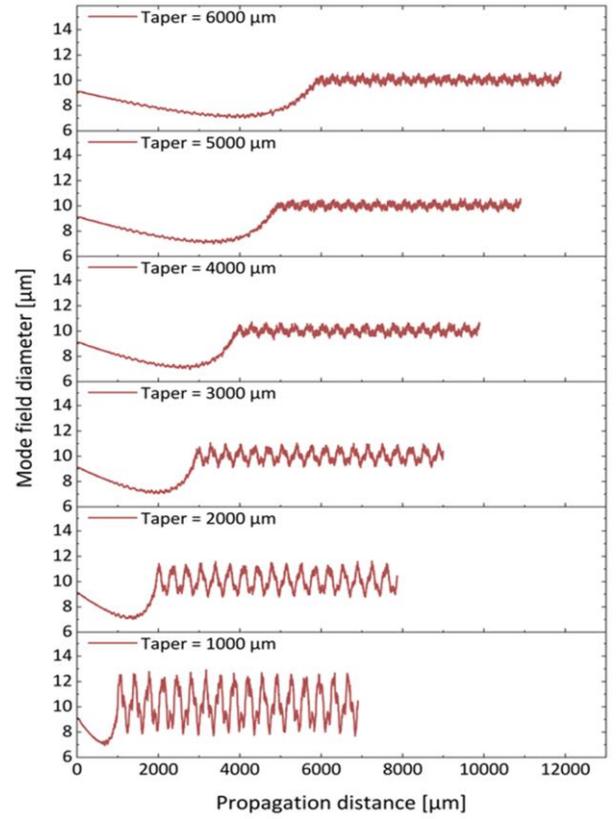

**Fig. 3.** The microlens on the tapered SMF: (a) scheme modeled in the BeamPROP software, (b) simulated light propagation profiles with focal length ($f_L$) and radius ($R$), (c) example of a fabricated structure with taper length of 40 μm and ball size of 88 μm.

Fig. 3 (b) shows an example simulation of the modeled structure presented in Fig. 3 (a). The tapered diameter was set to 40 μm and the diameter of the microlens was set to 80 μm. These input parameters resulted in achieving the beam radius value equal to 6.3 μm and the focal equal to 21 μm. In Fig. 3 (c) we present an LDS camera view of the fabricated structure, where the tapered diameter was equal to 40 μm and the diameter of the microlens was equal to 88 μm. In the fabrication process, the SMF was firstly tapered down in the LDS system to the desired diameter with the taper length of 5000 μm (by setting proper parameters of arc discharge power and pulling speed). During simulations, this value appeared to be the minimum desired value to achieve stable mode field diameter (around 10 μm) in the tapered waist segment before a microlens, which is shown in Fig. 4. Next, the taper waist section was cleaved to a length of approx. 1000 μm. The length of this segment is not crucial, but it fits the fabrication process perfectly and the mode field diameter does not change throughout this segment, which also can be seen in Fig. 4. Finally, a ball was created on the tip by applying a high-power discharge. Similarly to the previous case, the diameter of the ball can be easily controlled by adjusting the arc power of the ROF.

**Fig. 4.** Simulations of mode field diameter through the tapered SM fiber with the taper length range from 1000 μm to 6000 μm. The taper waist section is equal to 6000 μm in all cases.

## III. CHARACTERIZATION OF FABRICATED MICROLENSES

The experimental setup for characterizing the fabricated microlenses is shown in Fig. 5. The beam from the fabricated structures was directed to the beam profiler (BP209IR1/M, Thorlabs) through the beam expander (x20), created by two lenses ($L_1$ – C260TMD-C Thorlabs, focal length = 15.29 mm; $L_2$ – LA1484-C-ML Thorlabs, focal length = 300 mm).

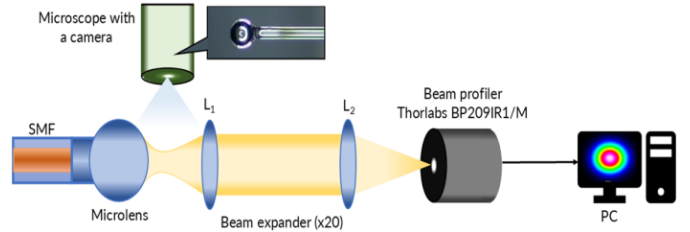

**Fig. 5.** Microlens characterization setup.

To measure the focal length of the fabricated structures, first, we set the cleaved SM fiber at a focal point (reference point) of the first beam expander lens. As an additional result, we measured its mode field diameter on the beam profiler and marked the position in microscope software. Next, we placed a microlens structure exactly in the same position as the SM fiber end face and then looked for the focal spot of the fabricated lens. The position of a structure was again marked, so the distance between the reference measurement (the focal point of cleaved SM fiber) and the actual position of the microlens was determined as the focal length. As the light source, we used a 1550 nm laser diode pigtailed with SMF.

*A. Ball-type microlens – experimental results*

The ball-type microlens design has two degrees of freedom: the length of the coreless fiber section, and the size (diameter) of the ball. In the experiment, we fabricated a series of structures with CPF lengths varying from 100 to 1100 μm, and ball diameters from 220 to 320 μm. Figure 6 shows the beam parameters obtained for different ball diameters as a function of the CPF length.

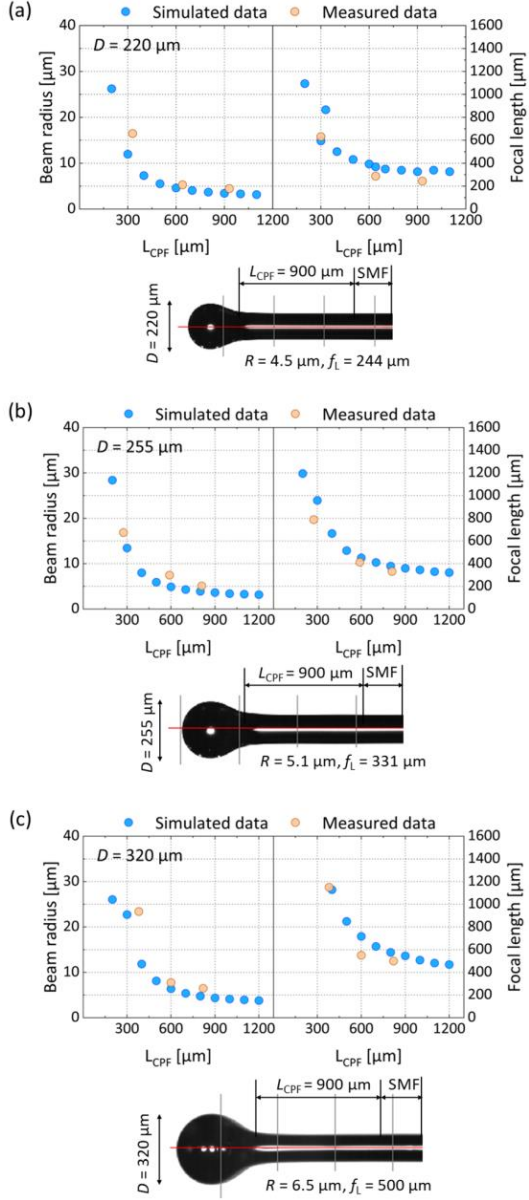

**Fig. 6.** Simulation and measurement results for the microlenses on the end of a fiber of beam radius and focal length as a function of the length of the coreless passive fiber and LDS camera view with: (a) microlens $D$ = 220 μm, (b) microlens $D$ = 255 μm, (c) microlens $D$ = 320 μm.

The range of possible focal lengths is very broad, ranging from 300 μm even up to almost 1200 μm. Depending on the parameters of the microlens, the spot radius could be as small as 4.5 μm (for a ball of $D$=220 μm and $L_{CPF}$ of 930 μm). The measured beam radius and focal length values agree with the simulations. Using the LDS system, we can easily control processing parameters and create microlenses with a large range of output parameters that can be adjusted for specific applications.

Figure 7 shows the measured beam profiles (i.e., the waist along the propagation axis over a distance of 1.5 mm) from a selected microlens with a diameter of 255 μm and three $L_{CPF}$ (300, 600, 900 μm).

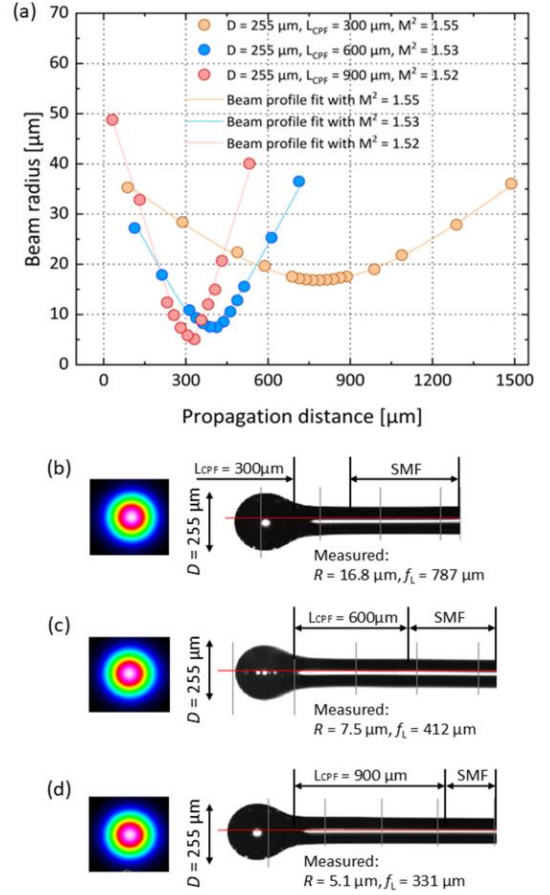

**Fig. 7.** Measurements of the beam radius and focal length of the ball-type microlenses: (a) measured beam profiles of fabricated structures, and beam profile and LDS camera view of the structure with (b) CPF length = 300 μm, (c) CPF length = 600 μm, (d) CPF length = 900 μm.

It can be seen from the measurements, that the bigger the microlens diameter, the larger the beam radius and the focal length (Fig. 6). On the contrary, the longer the length of the CPF, the smaller the beam radius, and the shorter the focal length. In all cases, the beams maintained an excellent beam shape (measured 2D intensity profiles shown in Fig. 7), and ellipticity of 99%. A fitting was applied to the beam profile data using the standard formula for the beam waist $w(z)$, the Rayleigh length $z_R$, and the $M^2$ factor [38]:

$$w(z) = w_0 \sqrt{1 + (\frac{z-z_0}{z_R})^2} \quad (1)$$

Where $w_0$ is the minimum beam waist, and $z_0$ is the location of the beam waist. The Rayleigh length $z_R$ is related to the $M^2$ parameter by the equation:

$$z_R = \frac{\pi w_0^2}{\lambda} M^2 \qquad (2)$$

where λ is the wavelength of the beam equal to 1550 nm. We note that the calculated $M^2$ parameters were not meant to verify the beam quality, since the performed measurements do not fulfill the ISO Standard 11146 (i.e., the number of points in the proximity of the waist, etc.); we only found the proper $M^2$ values to fit measurement results.

*B. Fiber tip microlens – experimental results*

For the fiber tip, we estimated that the taper sections are going down to the minimum diameter of the SMF equal to around 1 μm (Fig. 2). The experiment's results using the sharp and curved tips are shown in the figure below.

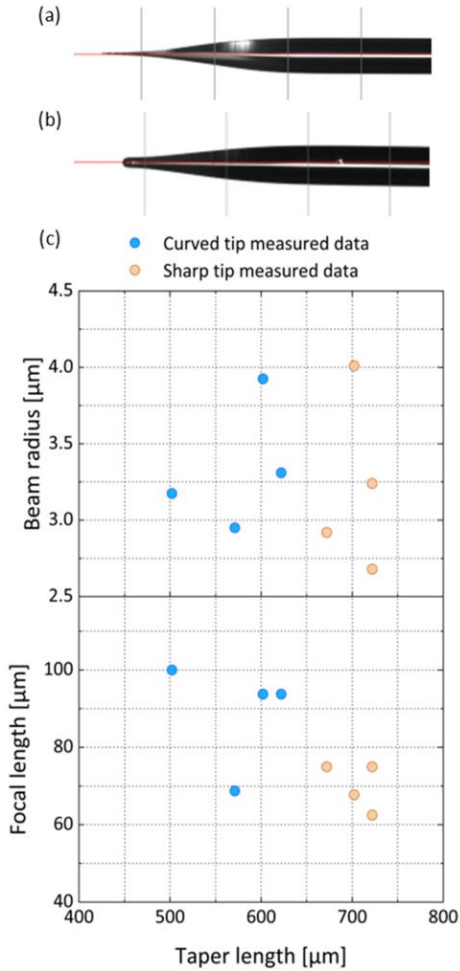

**Fig. 8.** Measurement results for the fiber tip microlenses: (a) an example of the sharp tip LDS camera view, (b) an example of the curved tip LDS camera view, (c) the beam radius and focal length as a function of the taper length.

For the fiber tips (curved and sharp) we observed a change in the beam radius from 2.7 μm to 4 μm with a focal length range of 62 μm to 100 μm. Taper lengths were between 502 – 722 μm. It stands out that the focal length for the curved tips has a larger range of focal length (up to 100 μm instead of 75 μm).

Another advantage that this kind of shape gives us is that we can easily fix the structure of the sharp tip by curving it, without much loss of the beam radius and focal length.

Figures 9 and 10 present the measured beam profile (i.e., the waist along the propagation axis over a distance of 0.2 mm) from a selected fiber tip with a length from 502 μm to 622 μm in the case of curved tips and a length from 672 μm to 722 μm in the case of sharp tips. It can be seen that curved tips result in a longer focal length and larger spot size in the waist. In all cases, the beams maintained an excellent beam shape (measured 2D intensity profiles shown in Fig. 7), and ellipticity of 99%.

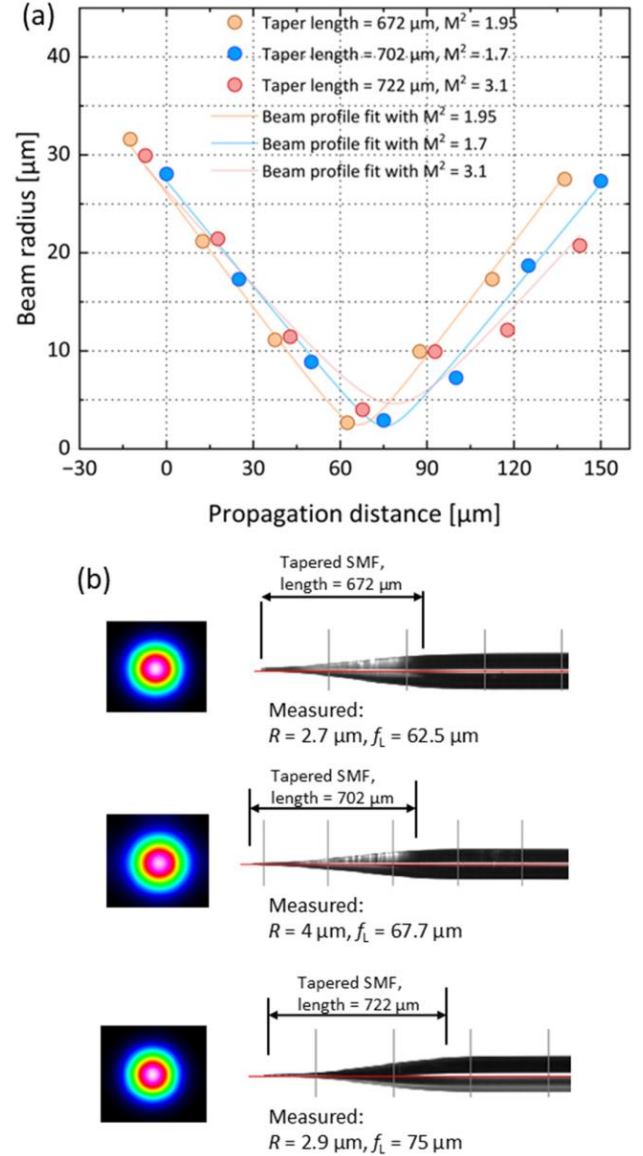

**Fig. 9**. Measurements of the beam radius and focal length of the optical fiber sharp tips: (a) measured beam profiles along the propagation axis, (b) beam profiles and LDS camera views of the structures with tapered SMF length = 672 μm - 722 μm.

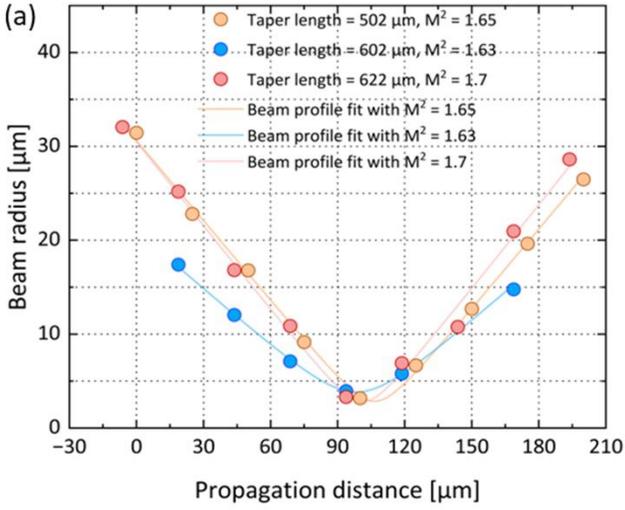

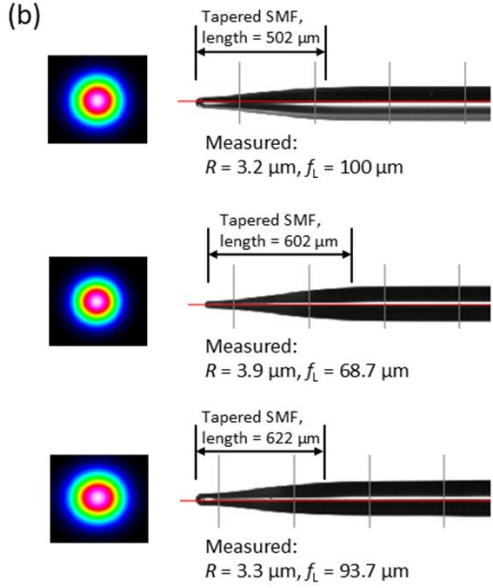

**Fig. 10.** Measurements of the beam radius and focal length of the optical fiber curved tips: (a) measured beam profiles along the propagation axis, (b) beam profiles and LDS camera views of the structures with tapered SMF length = 502 μm - 622 μm.

*C. Ball-type microlens on a tapered fiber – experimental results*

We noticed similar dependencies in the case of microlens on the tapered fiber end. It appears that, again, the beam radius and the focal length are increasing with the microlens diameter. On the other hand the smaller the diameter of the SMF taper waist ($D_{TAPER}$), the smaller the beam radius, and the shorter the focal length. In Fig. 11, we compared experimental results with simulations. We observed a mismatch between simulations and measured data, which is most likely caused by the imperfection of the BPM method, which may not be the optimal tool to simulate that kind of complex structure.

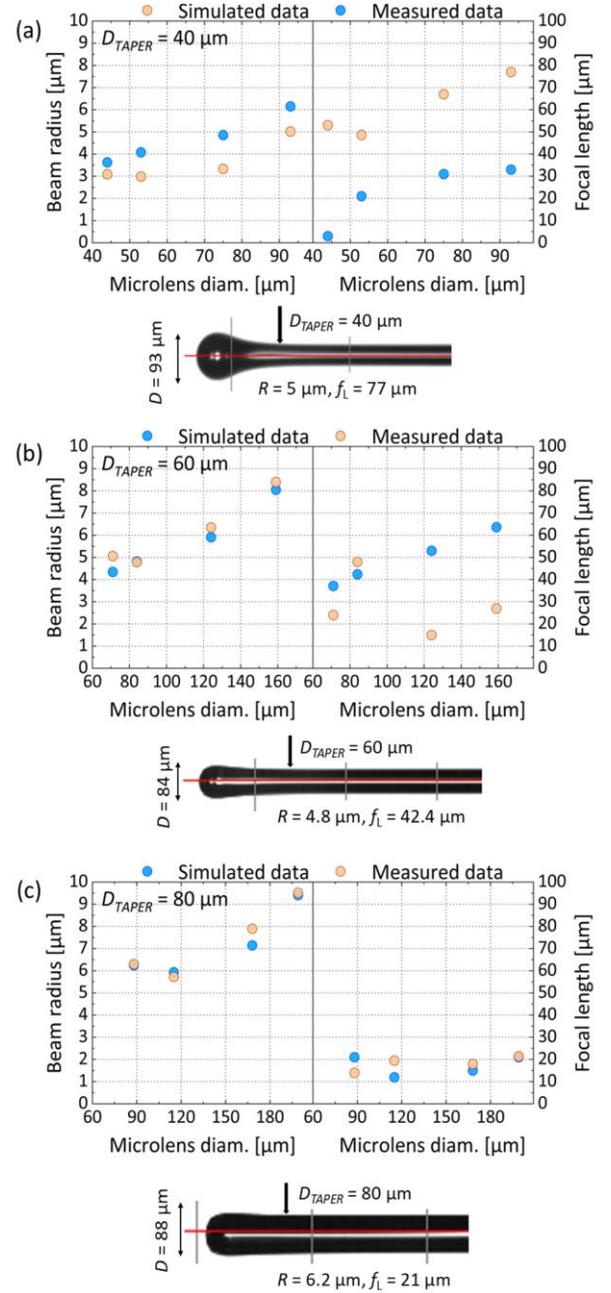

**Fig. 11**. Simulation and measurement results for the microlenses on a tapered fiber of beam radius and focal length as a function of the microlens diameter and LDS camera view for: (a) $D_{TAPER}$ = 40 μm, (b) $D_{TAPER}$ = 60 μm, (c) $D_{TAPER}$ = 80 μm.

Figure 12 shows the measured beam profile (i.e., the waist along the propagation axis over a distance of 0.1 mm) from selected microlens with a diameter range from 71 μm to 124 μm and constant $D_{TAPER}$ diameters (60 μm). It can be seen that a larger microlens diameter results in a longer focal length and bigger spot size at the waist. In all cases, the beams maintained an excellent beam shape (measured 2D intensity profiles shown in Fig. 11), and ellipticity of 99%.

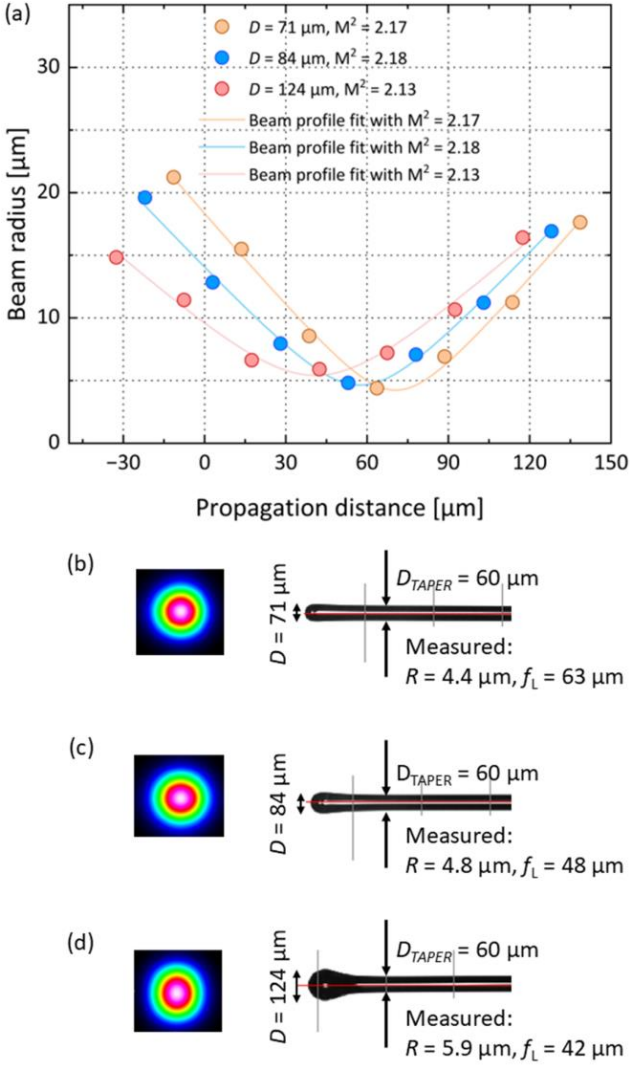

**Fig. 12.** Measurements of the beam radius and focal length of the microlenses on a tapered fiber: (a) measured beam profiles along the propagation axis of various microlenses diameter and with a constant $D_{TAPER}$ value, (b-d) beam profile and LDS camera view for microlens $D = 71$ µm, $D = 84$ µm, and $D = 124$ µm, respectively.

## IV. DISCUSSION AND CONCLUSIONS

The results of our research confirm that it is possible to fabricate various optical fiber microlenses using the LDS system. Several factors can cause inaccuracies between simulations and experimental results in our study. Firstly, simulated models were simplified, and conditions during simulations were ideal, which does not fully reproduce the complexities of real-world conditions. In this study, we used the BPM to simulate the optical behavior of fabricated microlenses. While BPM is advantageous for its simplicity and computational efficiency, it has its limitations. BPM relies on the slowly varying envelope approximation, which assumes that the field changes gradually along the propagation direction. This approximation is less accurate for structures with sharp discontinuities, rapid variations in refractive index, or strong diffraction effects. BPM is also primarily suited for paraxial beams, where the angle of propagation relative to the optical axis is small. These limitations reduce the accuracy of BPM in modeling non-paraxial fields or in situations involving significant reflection and refraction [39]. In contrast, the Finite-Difference Time-Domain (FDTD) method offers superior accuracy for a wide range of electromagnetic simulations. FDTD does not rely on the slowly varying envelope approximation and can model complex, time-dependent interactions of light with structures. Its time-domain approach allows for the detailed analysis of transient behaviors and wideband responses, making it particularly suitable for simulating photonic devices with sharp features and non-linear materials. FDTD can handle arbitrary geometries and material inhomogeneities, providing a more realistic representation of physical phenomena [30-32]. Our experimental results showed good agreement with simulations for certain parameters, demonstrating the effectiveness of the LDS system in fabricating microlenses. The system's ability to manually adjust the geometrical parameters of fabricated structures and the short time of the fabrication process allows for precise customization of the optical properties, such as beam radius and focal length. Unlike conventional telecommunication splicers, the LDS system has three electrodes forming the ROF, which allows uniform heating of the structure. This fact, combined with reduced shaping time, is crucial in reducing the influence of gravity or vibration [33]. Thanks to that, we successfully fabricated various types of microlenses, including ball-type microlenses and fiber tips, achieving a high degree of control over their geometric and output light beam parameters. In the table below, we present a summary of achieved beam radius ranges, focal length ranges, ellipticity of beams, and $M^2$ factor.

TABLE I
SUMMARY OF ACHIEVED RESULTS

| Structure type | $R$ range [µm] | $f_L$ range [µm] | Ellipt. [%] | $M^2$ range |
|---|---|---|---|---|
| **Ball-type microlens** | 5 – 23.5 | 331 – 787 | 99 | 1.52 – 1.55 |
| **Sharp tip** | 2.7 – 4 | 62.5 – 75 | 99 | 1.7 – 3.1 |
| **Curved tip** | 2.9 – 3.9 | 69 – 100 | 99 | 1.63 – 1.7 |
| **Ball-type microlens on a tapered fiber** | 3 – 9.9 | 12 – 77 | 99 | 2.13 – 2.18 |

ACKNOWLEDGMENT

The authors thank Alicja Kwaśny for the help during data processing.

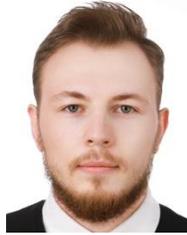

**Szymon Matczak** received his B.Sc. degree in Mechatronics, and M.Sc. in Electronics and telecommunications at Wrocław University of Science and Technology in 2021 and 2022, respectively.

He is currently pursuing a Ph.D. in designing and manufacturing fiber optics components for lasers and optical amplifiers at the Faculty of Electronics, Photonics, and Microsystems, Wroclaw University of Science and Technology.

Mr. Matczak's research interests include fiber optics, laser systems, and amplifiers. He is a student member of Optica.

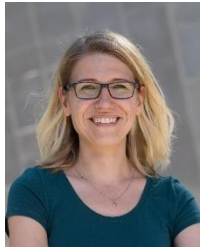

**Dorota Stachowiak** was born in Nowa Sól, Poland, on November 25, 1988. She earned her PhD in 2017, at the Wroclaw University of Science and Technology (WrUST), Faculty of Electronics (Wrocław, Poland). She received a B.S. degree in optoelectronics and an M.S. degree in acoustics from the WrUST, Wroclaw, Poland, in 2011 and 2012, respectively.

In 2012, she became a member of the Laser and Fiber Electronics Group, at WrUST, where she started her Ph.D. studies. From 2015 to 2019, she worked at the PORT Polish Center for Technology Development Wroclaw, Poland), in the Laser Sensing Laboratory as a process engineer (2015-2017) and research engineer (2017-2019). In 2019, she returned to research work at WrUST and is currently working as an assistant professor in the research group of the Laser and Fiber Electronics Group at the Faculty of Electronics, Photonics and Microsystems.

Her research deals with a fiber laser and amplifier technology, especially with the fabrication of components from which these systems are built, such as power combiners, mode field adaptors, couplers, and others, like e.g. fiber optic lenses.

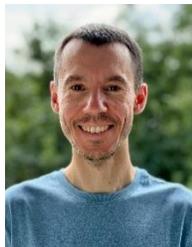

**Grzegorz Sobon** was born in Wrocław, Poland, on January 23, 1986. He received his doctoral degree and habilitation at Wroclaw University of Science and Technology (WrUST) in 2013 and 2018, respectively.

Since 2010, he is employed at WrUST. He was involved in over 30 research projects, of which in 7 as principal investigator. Since 2019, he is employed as associate professor.

His research interests focus on ultrafast fiber lasers, nonlinear fiber optics, optical frequency combs and laser spectroscopy.